\documentclass[12pt]{article}
\usepackage{amsmath}
\usepackage{amssymb}
\def\ga{\gamma}

\def\Ga{\Gamma}

\newcommand{\beq}{\begin{equation}}
\newcommand{\eeq}{\end{equation}}
\newcommand{\bea}{\begin{eqnarray*}}
\newcommand{\eea}{\end{eqnarray*}}
\newcommand{\beaq}{\begin{eqnarray}}
\newcommand{\eeaq}{\end{eqnarray}}
\makeatletter
\def\numberbysection{\@addtoreset{equation}{section}
\def\theequation{\thesection.\arabic{equation}}}
\makeatother

\numberbysection
\begin{document}
\centerline{\Large\bf One-Point Functions of}
\vskip .5cm
\centerline{\Large\bf $N=2$ Super-Liouville Theory with Boundary}
\vskip 2cm
\centerline{\large Changrim Ahn,\footnote{ahn@ewha.ac.kr}
Marian Stanishkov,\footnote{stanishkov@dante.ewha.ac.kr;
On leave of absence from Institute of Nuclear Research and Nuclear
Energy, Sofia, Bulgaria}
and Masayoshi Yamamoto\footnote{yamamoto@dante.ewha.ac.kr}}
\vskip 1cm
\centerline{\it Department of Physics}
\centerline{\it Ewha Womans University}
\centerline{\it Seoul 120-750, Korea}
\centerline{\small PACS: 11.25.Hf, 11.55.Ds}
\vskip 1cm
\centerline{\bf Abstract}
We derive one-point functions of the $N=2$ super-Liouville 
theory on a half line using the modular transformations of the characters
in terms of the bulk and boundary cosmological constants.
We also show that these results are consistent with
conformal bootstrap equations which are based on the bulk and 
boundary actions.
We provide various independent checks for our results.

%\newpage

\section{Introduction}
Two-dimensional Liouville field theory (LFT) has been studied 
for its relevance with non-critical string theories and two-dimensional
quantum gravity \cite{LFT,CurTho}.
This theory has been extended to the supersymmetric 
Liouville field theories (SLFTs) which can describe the non-critical 
superstring theories.
The LFT and SLFTs are irrational conformal field theories (CFTs) 
which have continuously infinite number of primary fields. 
It is very important to develope a CFT formalism which can apply to
these irrational CFTs.
There have been a lot of developments in this direction.
Various methods have been proposed \cite{Gervais} to derive structure
constants and reflection amplitudes, which are basic building
blocks to complete the conformal bootstrap \cite{Teschner,ZamZam}.
These have been extended to the $N=1$ SLFT in \cite{RasSta,Poghossian}.

More challenging problem is to extend these formalisms to the CFTs
with a boundary with a conformal boundary condition (BC).
Cardy showed that the conformal BCs
can be associated with the primary fields in terms of
modular $S$-matrix elements for rational CFTs \cite{Cardy}.
It has been an active issue to extend Cardy formalism 
to the irrational CFTs.
For example, the conformally invariant boundary states can 
be related to D-branes in the context of string theories 
\cite{string,DKKMMS,MMV}.
An important progress in this direction has been made in \cite{FZZ}.
With a boundary action which preserves the conformal symmetry, 
one-point function of a bulk operator and two-point correlation 
functions of boundary operators have been computed by 
conformal bootstrap method which extends the
functional relation method developed for the bulk theory \cite{Teschner}.
A similar treatment of the LFT defined in the classical Lobachevskiy
plane, namely the pseudosphere has been made in \cite{ZZ}.
This approach is generalized to the $N=1$ SLFT for the one-point functions 
\cite{ARS} and the boundary two-point functions \cite{FH}.

In this paper, we extend this approach to the $N=2$ SLFT.
This theory is of a particular interest in the string community
for rich properties \cite{n2slft}.
In spite of the extended symmetry, 
it turns out that exact correlation functions of 
the $N=2$ SLFT are much more difficult to derive than previous cases.
The main reason is that the $N=2$ SLFT has no strong-weak coupling duality.
The invariance of the LFT and $N=1$ SLFT under $b\to 1/b$ is realized
when the background charge changes to $1/b+b$ from its classical value
of $1/b$ after quantum corrections \cite{CurTho,ORaf}.
All the physical quantities like the correlations functions depend
on the coupling constant through this combination.
This invariance maintains an equivalence between a weak $b<<1$
and strong $b>>1$ coupling limits.
This duality as well as the functional relations based on the 
conformal bootstrap methods are essential ingredients to obtain
exact correlation functions uniquely for the LFT \cite{ZamZam}, 
the $N=1$ SLFT \cite{RasSta}, and their boundary extensions 
\cite{FZZ,ZZ,ARS,FH}. 

Differently from its simpler relatives, the $N=2$ SLFT is \emph{not} 
renormalized and no duality appears.
This nonrenormalization is a general aspect of $N=2$ supersymmetric 
quantum field theories in two-dimensional space-time.
Without the duality, the functional relations satisfied by the 
correlation functions can not be solved uniquely.
In \cite{AKRS}, an $N=2$ super-CFT has been proposed as a dual
theory to the $N=2$ SLFT under a transformation $b\to 1/b$..
Based on this conjecture, the bulk two-point functions, or ``reflection
amplitudes'', of both Neveu-Schwarz (NS) and Ramond (R) sectors
have been computed and various independent checks have been made.

Computing the one-point functions of the $N=2$ SLFT is more complicated.
The standard approach for one-point functions initiated 
by \cite{FZZ,ZZ} and followed by \cite{FZZ,ARS,FH} is
the conformal bootstrap method which can generate 
functional relations using the conformally invariant boundary actions
as boundary screening operators. 
The boundary action of the $N=2$ SLFT has been recently 
derived in \cite{AY}.
However, the lack of the duality prohibits this approach here too.
Namely, the $N=2$ SLFT with this boundary action is not self-dual either
and one needs to know the boundary action of the dual $N=2$ theory.
Without this, one can not solve the functional relations uniquely.
Due to the nonlocality of the bulk action of the dual $N=2$ theory 
\cite{AKRS}, the method used in the $N=2$ SLFT \cite{AY} seems 
not be applicable. 
We need a different approach.
In the previous works \cite{ZZ,ARS,FH}, 
one-point functions have been obtained from the conformal bootstrap 
methods and confirmed by modular transformation properties using 
a known relationship to the conformal boundary states \cite{CarLew}.
In this paper we reverse the steps; we first derive the one-point
functions from the modular transformation properties. 
Then, we relate them to the bulk and boundary actions of the 
$N=2$ SLFT and its dual theory by conformal bootstrap methods.
Although the functional relations obtained in this way is not complete
due to the limitation we already mentioned, they can give
essential informations which the modular transformations can not provide.

There appeared already a paper which used the modular transformation method
to find the one-point functions and associated boundary states \cite{MMV}.
However, this paper considers only a special value of coupling constant and 
only a vacuum BC. 
Moreover, a direct relation to the $N=2$ SLFT is missing.
Our results include general one-point functions 
for general BCs parametrized by a continuous parameter.
We also provide various consistency checks.
Using the one-point functions, we rederive the bulk reflection 
amplitudes and compare them with those
derived  independently \cite{BasFat,AKRS}.
Also we provide semiclassical checks.
Furthermore, as previously described, we provide conformal 
boostrap analysis based on the $N=2$ SLFT and its dual theories and
confirme the one-point functions obtained from the modular transformations
are consistent with the bulk and boundary actions.
As byproducts, we obtain a relation between the continuous BC  parameter 
and the boundary cosmological constants of the two dual theories.

This paper is organized as follows. In sect.2 we introduce the bulk 
and boundary actions of the $N=2$ SLFT along with notations.
Then, we derive the characters and their modular transformations in sect.3.
In sect.4, we present our main results, the one-point 
functions and functional relations from the conformal bootstrap 
along with several consistency checks.
We discuss a relation to boundary two-point functions and some concluding
remarks in sect.5.

%\newpage
\section{$N=2$ Super-Liouville Theory}

The action of the $N=2$ SLFT with the boundary is given by
\beaq
S&=&\int d^2z\Bigg[\frac{1}{2\pi}\left(\partial\phi^-\bar{\partial}\phi^+
+\partial\phi^+\bar{\partial}\phi^-
+\psi^-\bar{\partial}\psi^++\psi^+\bar{\partial}\psi^-
+\bar{\psi}^-\partial\bar{\psi}^++\bar{\psi}^+\partial\bar{\psi}^-\right)
\nonumber\\
&&+i\mu b^2\psi^-\bar{\psi}^-e^{b\phi^+}
+i\mu b^2\psi^+\bar{\psi}^+e^{b\phi^-}
+\pi\mu^2 b^2e^{b(\phi^++\phi^-)}\Bigg]+S_B,
\label{N2L}
\eeaq
where the boundary action is derived to be \cite{AY}
\beaq
S_B&=&\int_{-\infty}^{\infty}dx\Bigg[-\frac{i}{4\pi}
(\bar{\psi}^+\psi^-+\bar{\psi}^-\psi^+) +\frac{1}{2}a^-\partial_x a^+
\nonumber\\
&-&\frac{1}{2}e^{b\phi^+/2}\left(\mu_B a^+
+\frac{\mu b^2}{4\mu_B}a^-\right)(\psi^-+\bar{\psi}^-)
-\frac{1}{2}e^{b\phi^-/2}\left(\mu_B a^-
+\frac{\mu b^2}{4\mu_B} a^+\right)(\psi^++\bar{\psi}^+)\nonumber\\
&-&\frac{2}{b^2}\left(\mu_B^2+\frac{\mu^2b^4}{16\mu_B^2}\right)
e^{b(\phi^++\phi^-)/2}\Bigg].\label{baction}
\eeaq
As in the LFT and the $N=1$ SLFT, one should introduce a background charge
$1/b$ so that the interaction terms in Eq.(\ref{N2L}) become the screening
operators of the conformal field theory (CFT).
As mentioned earlier, this background charge is unrenormalized due 
to the $N=2$ supersymmetry and the $N=2$ SLFT is not self-dual.

The stress tensor $T$, the supercurrent $G^{\pm}$ and
the $U(1)$ current $J$ are given by
\beaq
&&T=-\partial\phi^-\partial\phi^+
-\frac{1}{2}(\psi^-\partial\psi^++\psi^+\partial\psi^-)
+{1\over{2b}}(\partial^2\phi^++\partial^2\phi^-),
\label{N2T}\\
&&G^{\pm}=\sqrt{2}i(\psi^{\pm}\partial\phi^{\pm}-{1\over{b}}
\partial\psi^{\pm}),\qquad
J=-\psi^-\psi^{+}+{1\over{b}}(\partial\phi^+-\partial\phi^-).
\label{N2J}
\eeaq
Using the mode expansions for the currents and their operator product
expansion, one can find the $N=2$ super-Virasoro algebra
\bea
\left[L_m,G^{\pm}_{r}\right]&=&\left({m\over{2}}-r\right)G^{\pm}_{m+r},
\qquad \left[J_n,G^{\pm}_{r}\right]=\pm G^{\pm}_{n+r},\\
\left\{G^{+}_{r},G^{-}_{s}\right\}&=&2L_{r+s}+(r-s)J_{r+s}
+{c\over{3}}\left(r^2-{1\over{4}}\right)\delta_{r+s},\quad
\left\{G^{\pm}_{r},G^{\pm}_{s}\right\}=0,\\
\left[L_m,J_{n}\right]&=&-nJ_{m+n},\qquad
\left[J_m,J_{n}\right]={c\over{3}}m\delta_{m+n},
\eea
with the central charge
\beq
c=3+6/b^2.
\eeq
Due to anti-periodicity for the (NS) sector, 
the fermionic modes are given by half-integers
while for the (R) sector they are integer modes due to the periodicity.

The primary fields of the $N=2$ SLFT are classified into the
(NS) and the (R) sectors and can be written
in terms of the component fields as follows \cite{Marian}:
\begin{equation}
N_{\alpha{\overline\alpha}}=
e^{\alpha\phi^{+}+{\overline\alpha}\phi^{-}},\qquad
R^{\pm}_{\alpha{\overline\alpha}}=
\sigma^{\pm}e^{\alpha\phi^{+}+{\overline\alpha}\phi^{-}},
\label{primary}
\end{equation}
where $\sigma^{\pm}$ is the spin operators.

The conformal dimensions and the $U(1)$ charges of the primary fields 
$N_{\alpha{\overline\alpha}}$ and $R^{\pm}_{\alpha{\overline\alpha}}$ 
can be obtained:
\begin{equation}
\Delta^{NS}_{\alpha{\overline\alpha}}=-\alpha{\overline\alpha} 
+{1\over{2b}}(\alpha+{\overline\alpha}),\qquad
\Delta^{R}_{\alpha{\overline\alpha}}=\Delta^{NS}_{\alpha{\overline\alpha}}
+{1\over{8}},
\label{delta}
\end{equation}
and 
\begin{equation}
\omega={1\over{b}}(\alpha-{\overline\alpha}),
\qquad
\omega^{\pm}=\omega \pm{1\over{2}}.
\label{u1charge}
\end{equation}
It is more convenient to use a `momentum' defined by
\beq
\alpha+{\overline\alpha}={1\over{b}}+2iP, \label{momentum}
\eeq
and the $U(1)$ charge $\omega$ instead of $\alpha,{\overline\alpha}$.
In terms of these, the conformal dimensions are given by 
\begin{equation}
\Delta^{NS}={1\over{4b^2}}+P^2+{b^2\omega^2\over{4}}.
\label{deltanew}
\end{equation}
From now on, we will denote a (NS) primary state by
$\vert[P,\omega]\rangle$ and a (R) state by 
$\vert[P,\omega,\epsilon]\rangle$ with $\epsilon=\pm 1$.

One can notice that the conformal dimenions and $U(1)$ charges
are invariant under
\beq
\alpha\to 1/b-{\overline\alpha},\qquad
{\overline\alpha}\to 1/b-\alpha. 
\label{reflection}
\eeq
This means $N_{\alpha{\overline\alpha}}$ should be identified
with $N_{1/b-\alpha,1/b-{\overline\alpha}}$ and similarly, 
for the (R) primary fields, up to normalization factors. 
In terms of the momentum parameter, this means an invariance under
$P\to-P$. 
In semiclassical picture where the primary fields can be described by
plane waves with momentum $P$ in the bosonic zero-mode space, 
this relation would imply that the wave with a momentum $P$ is
reflected off from the potential wall and changes the momentum to $-P$.
This qualitative description can be extended to the full quantum
region where the exact reflection amplitudes are defined and
computed using the functional relation methods. 
We will be back to this issue in sect.4.

%\newpage
\section{Characters and Modular Transformations}

A character is defined by the following trace over
all the conformal states built on a specific primary state:
\beq
\chi_h(q,y,t)=e^{2\pi ik t}{\rm Tr}\left[q^{L_0-c/24}y^{J_0}\right].
\eeq
Here $k$ is a fixed constant for a given CFT and we set
$k=1+2/b^2$ for the $N=2$ SLFT.
In terms of the modular parameters given by
\beq
q=e^{2\pi i\tau},\qquad y=e^{2\pi i\nu},\qquad
q'=e^{-2\pi i/\tau},\qquad y'=e^{2\pi i\nu/\tau}
\eeq
the modular group $SL(2,{\mathbf Z})$ is generated by the two
elements $T,S$
\beq
T:(\tau,\nu,t)\to(\tau+1,\nu,t),\qquad S:
(\tau,\nu,t)\to(-1/\tau,\nu/\tau,t-\nu^2/2\tau).
\label{modular}
\eeq
While the character transforms simply under $T$
\beq
\chi_h(\tau,\nu,t)=e^{2\pi i(h-c/24)}\chi_h(\tau,\nu,t),
\eeq
the characters transform under $S$ nontrivially and 
are expressed by the modular $S$-matrix:
\beq
\chi_h(-1/\tau,\nu/\tau,t-\nu^2/2\tau)=\sum_{h'}
{\mathbf S}_{hh'}\chi_{h'}(\tau,\nu,t).
\eeq

\subsection{$N=2$ SLFT Characters}

To compute the $N=2$ SLFT characters, one should classify all
the decendents by acting the super-Virasoro generators on a
highest weight state, excluding not independent states \cite{BFK}.
If we denote a (NS) primary field by the momentum $P$ and $U(1)$ 
charge $\omega$, the decendents are given by
\beq
\cdots L^{n_2}_{-2}L^{n_1}_{-1} \cdots J^{m_2}_{-2}J^{m_1}_{-1}
\cdots {G_{-3/2}^{+}}^{\epsilon^{+}_{3/2}}
{G_{-1/2}^{+}}^{\epsilon^{+}_{1/2}}
\cdots {G_{-3/2}^{-}}^{\epsilon^{-}_{3/2}}
{G_{-1/2}^{-}}^{\epsilon^{-}_{1/2}} \vert[P,\omega]\rangle
\label{decendents}
\eeq
where the exponents $n_i,m_i$ are arbitrary nonnegative integers 
and $\epsilon^{\pm}_r=0,1$ since ${G_{r}^{\pm}}^{2}=0$.

For generic values of $P,\omega$, the $N=2$ SLFT has no null states and
the characters can be obtained by simply summing the states.
Using the definition given above, the character is computed to be
\beaq
\chi^{NS}_{[P,\omega]}(q,y,t)&=&e^{2\pi i kt}q^{-1/8+P^2+b^2\omega^2/4}
y^{\omega} \prod_{n=1}^{\infty}{(1+yq^{n-1/2})
(1+y^{-1}q^{n-1/2})\over{(1-q^n)^2}}\label{nschi}\\
&=&e^{2\pi i kt} q^{P^2+b^2\omega^2/4} y^{\omega}
{\theta_{00}(q,y)\over{\eta(q)^3}},\nonumber
\eeaq
where we have introduced standard elliptic functions 
in the second line
\beaq
\eta(q)&=&q^{1/24}\prod_{n=1}^{\infty}(1-q^n),\\
\theta_{00}(q,y)&=&\prod_{n=1}^{\infty}\left[
(1-q^n)(1+yq^{n-1/2})(1+y^{-1}q^{n-1/2})\right].
\eeaq
The denominator of Eq.(\ref{nschi}) originates from the modes $L_{-n}$
and $J_{-n}$ and the numerators from $G^{\pm}_{-r}$.

For the conformal BCs of super-CFTs, one needs to consider 
characters and associated Ishibashi states of the 
$\widetilde{({\rm NS})}$ sectors \cite{Nepomechie}.
The $\widetilde{({\rm NS})}$ characters are defined by by
\beq
\chi^{\widetilde{NS}}_h(q,y,t)=
e^{2\pi i kt}{\rm Tr}\left[(-1)^Fq^{L_0-c/24}y^{J_0}\right].
\eeq
For a $N=2$ SLFT primary field with $[P,\omega]$, $(-1)^F$ term
contributes $-1$ for those decendents with odd number of $G^{\pm}_{-r}$.
This effect can be efficiently incorporated into the character formula
by shifting $y\to -y$ in the product. 
Therefore, the $\widetilde{({\rm NS})}$ character is given by
\beq
\chi^{\widetilde{NS}}_{[P,\omega]}(q,y,t)=e^{2\pi i kt}
q^{P^2+b^2\omega^2/4} y^{\omega} {\theta_{00}(q,-y)\over{\eta(q)^3}}.
\label{nstildechi}
\eeq

The characters of the (R) sector are rather different.
Decendents of a (R) primary field with $[P,\omega,\epsilon]$ whose conformal
dimension and charge are given in Eqs.(\ref{deltanew}) and (\ref{u1charge}) 
are constructed by acting $L_{-n}$'s, $J_{-n}$'s and $G^{\pm}_{-r}$'s.
While $n$ is any positive integer, one should be careful for $r$.
As noticed in \cite{BFK}, the (R) primary states satisfy
\beq
G_0^{\pm}\vert[P,\omega,\pm]\rangle=0.
\eeq  
Therefore, the (R) decendent module can include an extra state
$G_0^{\pm}\vert[P,\omega,\mp]\rangle$ and its decendents, respectively.
Including these at each value of $\epsilon$, one can find
the (R) character
\beaq
\chi^{R}_{[P,\omega,\epsilon]}(q,y,t)&=&
e^{2\pi i kt} q^{P^2+b^2\omega^2/4}
y^{\omega+\epsilon/2}(1+y^{-\epsilon}) \prod_{n=1}^{\infty}{(1+yq^{n})
(1+y^{-1}q^{n})\over{(1-q^n)^2}}\nonumber\\
&=& e^{2\pi i kt} q^{P^2+b^2\omega^2/4} y^{\omega}
{\theta_{10}(q,y)\over{\eta(q)^3}},\label{rchi}
\eeaq
where we introduce another elliptic function
\beq
\theta_{10}(q,y)=(y^{1/2}+y^{-1/2})q^{1/8}\prod_{n=1}^{\infty}\left[
(1-q^n)(1+yq^{n})(1+y^{-1}q^{n})\right].
\eeq

\subsection{Modular Transformations}

Here we consider only $S$ transformation $(q,y,t)\to (q',y',t')$
defined in Eq.(\ref{modular}) with $t'=t-\nu^2/2\tau$.
For irrational CFTs such as the SLFTs with infinite number of 
primary fields, the modular $S$-matrix will be indexed by 
continuous parameters and the summation will be replaced by integrations.

First, the modular transformation properties of the elliptic
functions are well-known:
\beq
{\theta_{00}(q',y')\over{\eta(q')^3}}=e^{\pi i\nu^2/\tau}
{i\over{\tau}}{\theta_{00}(q,y)\over{\eta(q)^3}},\quad
{\theta_{10}(q',y')\over{\eta(q')^3}}=e^{\pi i\nu^2/\tau}
{i\over{\tau}}{\theta_{00}(q,-y)\over{\eta(q)^3}}.
\eeq
Using these and Gaussian integrals, we have found the following 
modular transformations:
\beq
\chi^{NS}_{[P,\omega]}(q',y',t')=b\int_{-\infty}^{\infty}dP'
\int_{-\infty}^{\infty}d\omega' \cos(4\pi PP')e^{-\pi ib^2\omega\omega'}
\chi^{NS}_{[P',\omega']}(q,y,t),\label{nsmodtrans}
\eeq
\beq
\chi^{\widetilde{NS}}_{[P,\omega]}(q',y',t')=
b\int_{-\infty}^{\infty}dP'\int_{-\infty}^{\infty}
d\omega'\cos(4\pi PP')
e^{-\pi ib^2\omega\omega'}\chi^{R}_{[P',\omega',\epsilon]}(q,y,t),
\label{rmodtrans}
\eeq
\beq
\chi^{R}_{[P,\omega,\epsilon]}(q',y',t')=
b\int_{-\infty}^{\infty}dP'\int_{-\infty}^{\infty}d\omega'
\cos(4\pi PP')e^{-\pi ib^2\omega\omega'}
\chi^{\widetilde{NS}}_{[P',\omega']}(q,y,t).
\label{rmodtrans2}
\eeq

\subsection{Chiral Primary Fields}

An interesting class of $N=2$ SLFT primary fields is a chiral 
(and antichiral) primary field defined by
\beq
G^{+}_{-1/2}\vert [P,\omega]\rangle=0.\qquad 
\eeq
Antichiral fields are defined by $G^{-}_{-1/2}$. 
Since they are almost the same, we consider only the chiral fields.
If a primary field is chiral, then it should satisfy
\beq
G^{-}_{1/2}G^{+}_{-1/2}\vert [P,\omega]\rangle
=(2L_0-J_0)\vert [P,\omega]\rangle=0\quad\rightarrow\quad 2\Delta=\omega.
\eeq
This means 
\beq
{1\over{4b^2}}+P^2+{b^2{\omega}^2\over{4}}={\omega\over{2}}\quad
\to\quad P=i\left({b\omega\over{2}}-{1\over{2b}}\right).
\eeq
We denote $\vert\omega\rangle=\vert[P,\omega]\rangle$.
All the decendent states of a chiral primary
field including $G^{+}_{-1/2}$ mode must be truncated from the Hilbert space.
This means $\epsilon_{1/2}^{+}=0$ in Eq.(\ref{decendents}).

The character of a (NS) chiral primary field, then, can be 
written as Eq.(\ref{nschi}) except one difference that the term
$(1+yq^{n-1/2})$ starts from $n=2$ because the mode
$G^{+}_{-1/2}$ does not contribute.
This changes the character into
\beq
\chi^{NS}_{\vert\omega\rangle}(q,y,t)= 
e^{2\pi i kt} {q^{-1/4b^2} (yq^{1/2})^{\omega}
\over{1+yq^{1/2}}}{\theta_{00}(q,y)\over{\eta(q)^3}}.
\label{nschch}
\eeq

Using the method introduced in \cite{Miki}, one can derive the modular
transformation of the character as follows:
\beaq
\chi^{NS}_{\vert\omega\rangle}(q',y',t')&=&
\frac{b}{2}\int_{-\infty}^{\infty}dP' \int_{-\infty}^{\infty}d\omega' 
\Big[\frac{e^{-\pi ib^2\omega\omega'}\cosh[2\pi bP'(\omega-1-\frac{1}{b^2})]}
{2\cosh(\pi bP'+\frac{\pi ib^2\omega'}{2})
\cosh(\pi bP'-\frac{\pi ib^2\omega'}{2})} \nonumber\\
&+&\frac{e^{-\pi ib^2(\omega-1)\omega'}\cosh[2\pi bP'(\omega-\frac{1}{b^2})]}
{2\cosh(\pi bP'+\frac{\pi ib^2\omega'}{2})
\cosh(\pi bP'-\frac{\pi ib^2\omega'}{2})}\Big] \chi^{NS}_{[P',\omega']}(q,y,t)
\label{nsmodtransi}\\
&+&i\sum_{n\in {\mathbb Z}}\int_{\frac{1}{b^2}}^{1+\frac{1}{b^2}}d\omega' 
e^{-\pi i(2n\omega+\omega+\omega'-\frac{1}{b^2})}
q^{\frac{k}{2}n^2}y^{kn}
\chi^{NS}_{\vert\omega'\rangle}(q,yq^n,t).\nonumber
\eeaq
Similar formulae for the antichiral (NS) primary fields and 
(anti-) chiral fields of the (R) and other sectors can be obtained.

\subsection{Identity (Vacuum) Operator}

As we will see shortly, the identity operator plays 
a very important role in our derivation of the one-point functions.
Therefore, we need to derive the character of this.
The vacuum state $\vert 0\rangle=\vert[-i/2b,0]\rangle$ satisfies
\beq
G^{\pm}_{-1/2}\vert 0\rangle=0,\qquad
L_{-1}\vert 0\rangle=0.
\eeq
This means that allowed decendents are given by Eq.(\ref{decendents})
with restrictions that $n_1=\epsilon^{\pm}_{1/2}=0$.
After excluding these states, one can find the (NS) character
is given by
\beq
\chi^{NS}_{\vert 0\rangle}(q,y,t)=e^{2\pi i kt}{q^{-1/4b^2}(1-q)
\over{(1+yq^{1/2})(1+y^{-1}q^{1/2})}}{\theta_{00}(q,y)\over{\eta(q)^3}}.
\label{idch}
\eeq
It is obvious that the two factors in the denominators arise
in the same way as the chiral fields and the factor $1-q$ in
numerator comes from deducting the null state at level $1$.
If expanding the ``specialized'' character in a power series of $q$,
we obtain
\beq
\chi^{NS}_{\vert 0\rangle}(q,1,0)=
1+q+2q^{3/2}+3q^2+\ldots
\eeq
We can identify first few levels with explicit decendent states.
As expected, the level $1/2$ states are all truncated out and
there is only one state left at the level $1$, namely, 
$J_{-1}\vert 0\rangle$.
Two states at the level $3/2$ should be
$G^{\pm}_{-3/2}\vert 0\rangle$ and three states at the level 
$2$ are created by $L_{-2},J_{-2},J_{-1}^2$.

The modular transformation of Eq.(\ref{idch}) can be derived 
as before following \cite{Miki} 
\beaq
\chi^{NS}_{\vert 0\rangle}(q',y',t')&=&\int_{-\infty}^{\infty}dP
\int_{-\infty}^{\infty}d\omega {\mathbf S}_{NS}(P,\omega)
\chi^{NS}_{[P,\omega]}(q,y,t)
\label{idmodtrans}\\
&+&2\sum_{n\in{\mathbb Z}}\int_{\frac{1}{b^2}}^{1+\frac{1}{b^2}}
d\omega'\sin\left[\pi\left(\omega'-\frac{1}{b^2}\right)\right]
q^{\frac{k}{2}n^2}y^{kn}
\chi^{NS}_{\vert\omega'\rangle}(q,yq^n,t),\nonumber\\
{\mathbf S}_{NS}(P,\omega)&=&
{\sinh(2\pi bP)\sinh\left({2\pi P\over{b}}\right)\over{
2b^{-1}\cosh\left(\pi bP+{i\pi b^2\omega\over{2}}\right)
\cosh\left(\pi bP-{i\pi b^2\omega\over{2}}\right)}}.
\eeaq
The first terms in Eqs.(\ref{nsmodtransi},\ref{idmodtrans}),
which are of our main concern, can be also derived by expanding 
the denominators of Eqs.(\ref{nschch},\ref{idch}), applying 
Eq.(\ref{nsmodtrans}), and resumming formally the infinite terms.
However, this geometric sum can diverge and misses the 
contributions from the chiral primary characters.\footnote{
We thank the authors of \cite{EguSug} for pointing this out.}
As explained in \cite{EguSug}, these parts are necessary and have
physical meanings with respect to the spectral flows of the superconformal
field theories \cite{EguTao}.

We will need to know the modular transformation of the identity operator 
in the ($\widetilde{\rm NS}$) sector in the next section.
For this, the Hilbert space of the conformal tower of the
identity operator is the same as the (NS) sector.
As explained previously, the basic difference in this character
arises from the $(-1)^F$ which changes $y\to -y$ in effect.
Therefore, the character is given by
\beq
\chi^{\widetilde{NS}}_{\vert 0\rangle}(q,y,t)=e^{2\pi i kt}{q^{-1/4b^2}(1-q)
\over{(1-yq^{1/2})(1-y^{-1}q^{1/2})}}{\theta_{00}(q,-y)\over{\eta(q)^3}}.
\label{idtch}
\eeq
Using the previous method \cite{Miki}, 
one can find the modular transformation 
\beaq
\chi^{\widetilde{NS}}_{\vert 0\rangle}(q',y',t')&=&\int_{-\infty}^{\infty}dP
\int_{-\infty}^{\infty}d\omega {\mathbf S}_{R}(P,\omega)
\chi^{R}_{[P,\omega,\epsilon]}(q,y,t),\label{idtmodtrans}\\
&+&2\sum_{n\in{\mathbb Z+{1\over{2}}}}\int_{\frac{1}{b^2}}^{1+\frac{1}{b^2}}
d\omega'\sin\left[\pi\left(\omega'-\frac{1}{b^2}\right)\right]
q^{\frac{k}{2}(n^2-{1\over{4}})}y^{k(n-{1\over{2}})} 
\chi^{NS}_{\vert\omega'\rangle}(q,yq^n,t),\nonumber\\
{\mathbf S}_{R}(P,\omega)&=&
{\sinh(2\pi bP)\sinh\left({2\pi P\over{b}}\right)\over{
2b^{-1}\sinh\left(\pi bP+{i\pi b^2\omega\over{2}}\right)
\sinh\left(\pi bP-{i\pi b^2\omega\over{2}}\right)}}.
\eeaq

%\newpage
\section{One-Point Functions in the Presence of Boundary}

In this section, we compute exact one-point functions of the (NS) and (R)
bulk operators $N_{\alpha{\overline\alpha}}$ and 
$R^{\epsilon}_{\alpha{\overline\alpha}}$ of the $N=2$ 
SLFT with boundary.
The one-point functions are defined by
\beq
\langle N_{\alpha{\overline\alpha}}(\xi,\bar\xi)\rangle
={U^{NS}(\alpha,{\overline\alpha})\over 
{|\xi-\bar\xi|^{2\Delta^{NS}_{\alpha{\overline\alpha}}}}},\quad
{\rm and}\quad
\langle R^{\epsilon}_{\alpha{\overline\alpha}}(\xi,\bar\xi)\rangle 
={U^{R}(\alpha,{\overline\alpha})\over 
{|\xi-\bar\xi|^{2\Delta^R_{\alpha{\overline\alpha}}}}},
\eeq
with the conformal dimensions given in Eq.(\ref{delta}).
We will simply refer to the coefficients $U^{NS}(\alpha,{\overline\alpha})$ 
and $U^R(\alpha,{\overline\alpha})$ as the one-point functions.

\subsection{Vacuum Boundary Condition}

According to Cardy's formalism, one can associate
a conformal BC with each primary state \cite{Cardy}. 
For the $N=2$ SLFT, there will be infinite number of conformal BCs.
These BCs can be constructed by the fusion process and related to
the one-point functions.
Let us begin with the `vacuum' BC which corresponds to the identity
operator.
First we introduce an amplitude as an inner product between 
the Isibashi state of a primary state and the conformal boundary state\footnote{
We denote a conformal BC in `bold face' like ${\mathbf 0}$ and
a conformal boundary state like $\vert({\mathbf 0})\rangle$.}
\beq
\Psi_{\mathbf 0}^{NS}(P,\omega)=
\langle ({\mathbf 0})|[P,\omega]\rangle\rangle.
\eeq
From the modular transformation Eq.(\ref{idmodtrans}), the
amplitude satisfies the following relation:
\beq
\Psi_{\mathbf 0}^{NS}(P,\omega){\Psi_{\mathbf 0}^{NS}}^{\dagger}(P,\omega)
={\mathbf S}_{NS}(P,\omega).
\label{nsvac}
\eeq
Since ${\Psi_{\mathbf 0}^{NS}}^{\dagger}(P,\omega)
=\Psi_{\mathbf 0}^{NS}(-P,\omega)$, one can solve this up to some 
unknown constant as follows:
\beq
\Psi^{NS}_{\mathbf 0}(P,\omega)=\sqrt{{b^3\over{2}}}
\left(X_{NS}\right)^{{iP\over{b}}}
{\Gamma\left({1\over{2}}-ibP+{b^2\omega\over{2}}\right)
\Gamma\left({1\over{2}}-ibP-{b^2\omega\over{2}}\right)\over{
\Gamma\left(-{2iP\over{b}}\right)\Gamma\left(1-2ibP\right)}}.
\eeq
The unknown constant $X_{NS}$ does not depend on $P,\omega$
and can not be determined by the modular transformation alone.
We will derive this constant later in this section by comparing with
the bulk reflection amplitudes.

Similarly, for the (R) sector, we define the (R) amplitude by
\beq
\Psi_{\mathbf 0}^{R}(P,\omega)=
\langle ({\mathbf 0})|[P,\omega,\epsilon]\rangle\rangle
\eeq
which satisfies from Eq.(\ref{idtmodtrans}) 
\beq
\Psi_{\mathbf 0}^{R}(P,\omega){\Psi_{\mathbf 0}^{R}}^{\dagger}(P,\omega)
={\mathbf S}_{R}(P,\omega).
\label{rvac}
\eeq
The solution is up to a unknown constant:
\beq
\Psi^{R}_{\mathbf 0}(P,\omega)=-i\sqrt{{b^3\over{2}}}
\left(X_{R}\right)^{{iP\over{b}}}
{\Gamma\left(-ibP+{b^2\omega\over{2}}\right)
\Gamma\left(1-ibP-{b^2\omega\over{2}}\right)\over{
\Gamma\left(-{2iP\over{b}}\right)\Gamma\left(1-2ibP\right)}}.
\eeq
Again, the unknown constant $X_{R}$ will be fixed later.

\subsection{Continuous Boundary Condition}

Now we consider a continuous BC associated with a primary field.
This field should be (NS) and its $U(1)$ charge should be zero
because only the boundary neutral operators should appear.
So, we consider the character of a (NS) primary state
$\vert s\rangle\equiv\vert[s,0]\rangle$ and its modular transformation.
The parameter $s$ depends on the
boundary parameter $\mu_B$ in Eq.(\ref{baction}).
In this case Eq.(\ref{nsmodtrans}) becomes
\beq
\chi^{NS}_{\vert s\rangle}(q',y',t')= b\int_{-\infty}^{\infty}dP
\int_{-\infty}^{\infty}d\omega \cos(4\pi sP)
\chi^{NS}_{[P,\omega]}(q,y,t).\label{nsneumod}
\eeq
Now following previous analysis of the modular transformation,
this character should be written as
\beq
\chi^{NS}_{\vert s\rangle}(q',y',t')=
\int_{-\infty}^{\infty}dP \int_{-\infty}^{\infty}d\omega 
\Psi_{\mathbf s}^{NS}(P,\omega) {\Psi_{\mathbf 0}^{NS}}^{\dagger}(P,\omega)
\chi^{NS}_{[P,\omega]}(q,y,t).
\label{nsneumodi}
\eeq
Here we have defined an inner product between the conformal 
boundary state and an Ishibashi state
\beq
\Psi_{\mathbf s}^{NS}(P,\omega)
=\langle({\mathbf s})|[P,\omega]\rangle\rangle. 
\eeq

From Eqs.(\ref{nsneumod}) and (\ref{nsneumodi}), one can find 
\beq
\Psi^{NS}_{\mathbf s}(P,\omega){\Psi_{\mathbf 0}^{NS}}^{\dagger}(P,\omega)
=b\cos(4\pi s P). 
\eeq
Now acting $\Psi_{\mathbf 0}^{NS}(P,\omega)$ on this and using
Eq.(\ref{nsvac}), we obtain 
\beaq
\Psi^{NS}_{\mathbf s}(P,\omega)&=&b\Psi_{\mathbf 0}^{NS}(P,\omega)
{\cos(4\pi s P)\over{{\mathbf S}_{NS}(P,\omega)}}\nonumber\\
&=&\sqrt{2b^3}\left(X_{NS}\right)^{{iP\over{b}}}
{\Gamma\left(1+{2iP\over{b}}\right)\Gamma\left(2ibP\right)
\cos(4\pi s P)\over{\Gamma\left({1\over{2}}+ibP+{b^2\omega\over{2}}\right)
\Gamma\left({1\over{2}}+ibP-{b^2\omega\over{2}}\right)}}.
\label{ampns}
\eeaq

One can follow the same step for the (R) sector.
From Eq.(\ref{rmodtrans}) with $P=s,\omega=0$, one can find
\beq
\Psi^{R}_{\mathbf s}(P,\omega){\Psi_{\mathbf 0}^{R}}^{\dagger}(P,\omega)
=b\cos(4\pi s P),
\eeq
where 
\beq
\Psi_{\mathbf s}^{R}(P,\omega)=\langle({\mathbf s})|[P,\omega,\epsilon]
\rangle\rangle.
\eeq
Using Eq.(\ref{rvac}) on this, we can obtain
\beaq
\Psi^{R}_{\mathbf s}(P,\omega)&=&b\Psi_{\mathbf 0}^{R}(P,\omega)
{\cos(4\pi s P)\over{{\mathbf S}_{R}(P,\omega)}}\nonumber\\
&=&-i\sqrt{2b^3}\left(X_{R}\right)^{{iP\over{b}}}
{\Gamma\left(1+{2iP\over{b}}\right)\Gamma\left(2ibP\right)
\cos(4\pi s P)\over{\Gamma\left(1+ibP-{b^2\omega\over{2}}\right)
\Gamma\left(ibP+{b^2\omega\over{2}}\right)}}.
\label{ampr}
\eeaq

The amplitudes (\ref{ampns}) and (\ref{ampr}) we have obtained are
the one-point functions of the two sectors up to some normalization 
constants.
To fix these constants, we recall the relation proved in \cite{CarLew}
\beq
U_{\mathbf k}(\phi)={\langle({\mathbf k})|\phi\rangle\rangle\over{
\langle({\mathbf k})|0\rangle\rangle}}
\label{carlew}
\eeq
where ${\mathbf k}$ is a conformal BC, $\phi$ a primary field,
and $|\phi\rangle\rangle$, its Isibashi state.
For the $N=2$ SLFT, this relation means
\beq
U^{NS}_{\mathbf s}(P,\omega)
={\Psi_{\mathbf s}^{NS}(P,\omega)\over{
\Psi_{\mathbf s}^{NS}(-i/2b,0)}},\qquad
U^{R}_{\mathbf s}(P,\omega)
={\Psi_{\mathbf s}^{R}(P,\omega)\over{
\Psi_{\mathbf s}^{NS}(-i/2b,0)}}.
\eeq
From Eqs.(\ref{ampns}) and (\ref{ampr}) we can obtain
the one-point functions as follows:
\beaq
U^{NS}_{\mathbf s}(P,\omega)
&=&{\cal N}\left(X_{NS}\right)^{{iP\over{b}}}
{\Gamma\left(1+{2iP\over{b}}\right)\Gamma\left(2ibP\right)
\cos(4\pi s P)\over{\Gamma\left({1\over{2}}+ibP+{b^2\omega\over{2}}\right)
\Gamma\left({1\over{2}}+ibP-{b^2\omega\over{2}}\right)}},
\label{oneptns}\\
U^{R}_{\mathbf s}(P,\omega)&=&
{\cal N}\left(X_{R}\right)^{{iP\over{b}}}
{\Gamma\left(1+{2iP\over{b}}\right)\Gamma\left(2ibP\right)
\cos(4\pi s P)\over{\Gamma\left(1+ibP-{b^2\omega\over{2}}\right)
\Gamma\left(ibP+{b^2\omega\over{2}}\right)}}, \label{oneptr}
\eeaq
where the normalization coefficient ${\cal N}$ can be   
fixed by
\beq
U^{NS}_{\mathbf s}(-i/2b,0)=1\to
{\cal N}=\left[
\left(X_{NS}\right)^{1/2b^2}\Gamma(1+b^{-2})
\cosh\left({2\pi s\over{b}}\right)\right]^{-1}.
\eeq
Then, from $U^{R}_{\mathbf s}(-i/2b,0)=\langle\sigma^{\pm}\rangle$, 
one can find
\beq
\langle\sigma^{\pm}\rangle={2\over{\pi}}
\left({X_R\over{X_{NS}}}\right)^{1/2b^2}.
\eeq
The constants $X_{NS}$ and $X_R$ will be fixed shortly.

\subsection{Bulk Reflection Amplitudes}

The invariance of both conformal dimensions and $U(1)$ charges
under Eq.(\ref{reflection}) means that
$N_{1/b-{\overline\alpha},1/b-\alpha}$ should be identified
with $N_{\alpha{\overline\alpha}}$ and similarly for the (R) operators
up to normalization factors.
The reflection amplitudes are defined by
two-point functions of the same operators 
%\begin{eqnarray}
\beq
\langle N_{\alpha{\overline\alpha}}(z,{\overline z})
N_{\alpha{\overline\alpha}}(0,0)\rangle
={D^{NS}(\alpha,{\overline\alpha})\over{
|z|^{4\Delta^{NS}_{\alpha{\overline\alpha}}}}},\qquad
\langle R^{+}_{\alpha{\overline\alpha}}(z,{\overline z}) 
R^{-}_{\alpha{\overline\alpha}}(0,0)\rangle 
={D^{R}(\alpha,{\overline\alpha})\over{
|z|^{4\Delta^R_{\alpha{\overline\alpha}}}}}
\eeq
%\end{eqnarray}
with $\Delta^{NS}_{\alpha{\overline\alpha}},
\Delta^R_{\alpha{\overline\alpha}}$ given in Eq.(\ref{delta}).
In general, identification of the two fields gives a relation 
\beq
\langle N_{\alpha{\overline\alpha}}(z,{\overline z})\ldots\rangle
=D^{NS}(\alpha,{\overline\alpha})
\langle N_{{1\over{b}}-{\overline\alpha},{1\over{b}}-\alpha}
(z,{\overline z})\ldots\rangle
\label{refcorr}
\eeq
and similarly for the (R) sector.
Here the part $\ldots$ can be any products of the primary fields.

It turns out that the computation of these quantities is much
more complicated than that of the LFT or the $N=1$ SLFT case.
As we mentioned earlier, the reason is the lack of the self-duality.
In \cite{AKRS}, the reflection amplitudes of the primary fields 
with zero $U(1)$ charges have been derived based on a conjectured
$N=2$ super-CFT which is dual to the $N=2$ SLFT.
While these results are based on the conjecture, the resulting
reflection amplitudes have passed several consistency checks.
Moreover, these results are in exact agreement with
the reflection amplitudes which have been derived from
certain integrable field theory with two parameters proposed in
\cite{Fateev} which generates $N=2$ supersymmetry at special 
values of couplings \cite{BasFat}.
This agreement between two independent approaches strongly
supports the validity of the reflection amplitudes and the dual
action.

Here, we provide another independent derivation of the 
reflection amplitudes based on the one-point functions 
we have derived.
This computation will provide not only another confirmation of the results,
but also can be used to fix the undetermined constants.
The reflection relations among the correlation functions can be used
for a simpliest case, namely, the one-point functions.
In this case, the relation becomes
\beaq
\langle N_{\alpha{\overline\alpha}}(z,{\overline z})\rangle
&=&D^{NS}(\alpha,{\overline\alpha})
\langle N_{{1\over{b}}-{\overline\alpha},{1\over{b}}-\alpha}
(z,{\overline z})\rangle,\\
\langle R_{\alpha{\overline\alpha}}(z,{\overline z})\rangle
&=&D^{R}(\alpha,{\overline\alpha})
\langle R_{{1\over{b}}-{\overline\alpha},{1\over{b}}-\alpha}
(z,{\overline z})\rangle.
\eeaq
These lead to the following equations:
\beq
{U^{NS}_{\mathbf s}(P,\omega)\over{
U^{NS}_{\mathbf s}(-P,\omega)}}=D^{NS}(P,\omega),\qquad
{U^{R}_{\mathbf s}(P,\omega)\over{
U^{R}_{\mathbf s}(-P,\omega)}}=D^{R}(P,\omega).
\label{therelation}
\eeq

For the neutral sector $\omega=0$, the reflection amplitudes
has been derived in \cite{AKRS} 
\beaq
D^{NS}(P,0)&=&-\kappa^{-2iP/b}
{\Ga\left(1+{2iP\over{b}}\right)\over{\Ga\left(1-{2iP\over{b}}\right)}} 
{\Ga\left(1+iPb\right)\over{\Ga\left(1-iPb\right)}} 
{\Ga\left({1\over{2}}-iPb\right)\over{\Ga\left({1\over{2}}+iPb\right)}}, 
\label{eqfive}\\
D^{R}(P,0)&=&\kappa^{-2iP/b}
{\Ga\left(1+{2iP\over{b}}\right)\over{\Ga\left(1-{2iP\over{b}}\right)}} 
{\Ga\left(1-iPb\right)\over{\Ga\left(1+iPb\right)}} 
{\Ga\left({1\over{2}}+iPb\right)\over{\Ga\left({1\over{2}}-iPb\right)}}. 
\label{eqsix}
\eeaq
where 
\beq
\kappa ={\mu^2 \pi^2\over{2}}
\ga(-b^2-1)  \ga\left(1+{b^2\over{2}}\right)
\ga\left({b^2\over{2}}+{3\over{2}}\right),
\eeq
with $\gamma(x)=\Gamma(x)/\Gamma(1-x)$ and the bulk
cosmological constant $\mu$ in Eq.(\ref{N2L}).

Inserting $\omega=0$ and using (\ref{oneptns}) and (\ref{oneptr}), 
the reflection amplitudes in Eq.(\ref{therelation}) 
are indeed in exact agreement with Eqs.(\ref{eqfive}) and (\ref{eqsix})
if and only if we identify the constants
\beq
X_{NS}=X_{R}=\left[2^{2b^2}\kappa\right]^{-1}.
\eeq
This provides a nontrivial check and completes
our derivation for the one-point functions.
Furthermore, we can use Eq.(\ref{therelation}) to compute
the reflection amplitudes for $\omega\neq 0$ case 
\beq
D^{NS}(P,\omega)=(2^{2b^2}\kappa)^{-2iP/b}
{\Ga\left(1+{2iP\over{b}}\right)\over{\Ga\left(1-{2iP\over{b}}\right)}} 
{\Gamma\left(2ibP\right)\over{\Gamma\left(-2ibP\right)}}
{\Gamma\left({1\over{2}}-ibP+{b^2\omega\over{2}}\right)\over{
\Gamma\left({1\over{2}}+ibP+{b^2\omega\over{2}}\right)}}
{\Gamma\left({1\over{2}}-ibP-{b^2\omega\over{2}}\right)\over{
\Gamma\left({1\over{2}}+ibP-{b^2\omega\over{2}}\right)}}
\label{genrefns}
\eeq
and 
\beq
D^{R}(P,\omega)=(2^{2b^2}\kappa)^{-2iP/b}
{\Ga\left(1+{2iP\over{b}}\right)\over{\Ga\left(1-{2iP\over{b}}\right)}} 
{\Gamma\left(2ibP\right)\over{\Gamma\left(-2ibP\right)}}
{\Gamma\left(1-ibP-{b^2\omega\over{2}}\right)\over{
\Gamma\left(1+ibP-{b^2\omega\over{2}}\right)}}
{\Gamma\left(-ibP+{b^2\omega\over{2}}\right)\over{
\Gamma\left(ibP+{b^2\omega\over{2}}\right)}}
\label{genrefr}. 
\eeq
These results can be compared with those from the 
two-parameter family models \cite{BasFat}
and we checked that two independent results match exactly.

To complete our derivation of the one-point functions, we should 
relate the boundary parameter $s$ with the boundary cosmological
constant $\mu_B$ in Eq.(\ref{baction}).
In principle, one should do this by deriving the functional relations
following \cite{FZZ,ARS,FH}.
Solving these coupled equations, one can find the relation between the
boundary parameters.
However, without the boundary dual theory, this method does not
work well for the $N=2$ SLFT.
Instead, we analyze the pole structure of the one-point functions
and compare them with direct calculations using the bulk and boundary action.

For this, we consider one-point function of a neutral (NS) field 
$N_{\alpha\alpha}$
\beq
{\rm residue}\ {U^{NS}(\alpha)\over{{\cal N}}}
\bigg\vert_{\alpha=(b^{-1}-nb)/2}=
\langle e^{\alpha(\phi^{+}+\phi^{-})}\rangle=
\sum_{p,q}{1\over{p!q!}}\langle e^{\alpha(\phi^{+}+\phi^{-})}
V^p B^q\rangle_{0},
\eeq
where $V,B$ are the interaction terms in the bulk and boundary actions.
If we choose $n=1$ ($\alpha=1/2b-b/2$), all terms vanish except 
$p=0,q=2$ which can be easily computed:
\beaq
\big\langle e^{\alpha(\phi^{+}+\phi^{-})}\left(i/2\right) 
B^2\big\rangle_0
&=&-2{\overline\mu_B}^2 \int_{-\infty}^{\infty} dx_1 dx_2
\bigg|x_1-{i\over{2}}\bigg|^{-2\alpha b} 
\bigg|x_2-{i\over{2}}\bigg|^{- 2\alpha b} |x_1- x_2|^{-(1+b^2)}\nonumber \\
&=&8\pi{\overline\mu_B}^2\Gamma(-b^2)\gamma\left({1+b^2\over{2}}\right)
\sin\left(\pi{1+b^2\over{2}}\right),
\eeaq
with
\beq
{\overline\mu_B}^2=\mu_B^2+\frac{\mu^2b^4}{16\mu_B^2}.
\label{bcosmo}
\eeq
The residue of Eq.(\ref{oneptns}) at $\alpha={\overline\alpha}=1/2b-b/2$
becomes
\beq
{b\over{2}}(2^{2b^2}\kappa)^{1/2}{\Gamma(-b^2)\over{\Gamma
\left({1-b^2\over{2}}\right)^2}}\cosh(2\pi sb).
\eeq
Comparing these two, we find
\beq
{\overline\mu_B}^2 ={\mu b\over{32\pi}}\cosh(2\pi sb).
\label{boundparam}
\eeq

\subsection{Conformal Bootstrap Approach}

The procedure to derive the functional equations satisfied by
the one-point functions are identical to \cite{FZZ,ARS,FH}.
Consider two-point functions of neutral operators, 
\beq
G^{NS}_{\alpha}(\xi,\xi')=
\langle R^+_{-{1\over{2b}}}(\xi)N_\alpha(\xi')\rangle,\qquad
G^{R}_{\alpha}(\xi,\xi')=
\langle R^+_{-{1\over{2b}}}(\xi)R^-_\alpha(\xi')\rangle,
\eeq
where
$R^+_{-1/2b}$ is a degenerate (R) operator, whose OPEs are given by
\beaq
R^+_{-{1\over{2b}}}N_\alpha &=&\left[R^+_{\alpha-{1\over 2b}}\right]
+ C^{NS}(\alpha) \left[R^+_{\alpha+{1\over 2b}}\right],\\
R^+_{-{1\over{2b}}}R^-_\alpha &=&\left[N_{\alpha-{1\over 2b}}\right] 
+ C^{R}(\alpha) \left[N_{\alpha +{1\over 2b}}\right].
\eeaq
Here the bracket [\ldots] means the conformal tower of a given primary 
field and the structure constants have been computed in \cite{AKRS}
based on the dual $N=2$ SLFT:
\beaq
C^{NS}(\alpha) &=&{\tilde\mu}\pi\gamma\left(1+b^{-2}\right)
{\Gamma\left({2\alpha\over{b}}-{1\over{b^2}}\right)
\Gamma\left(1-{2\alpha\over{b}}\right)\over{
\Gamma\left(1-{2\alpha\over{b}}+{1\over{b^2}}\right)
\Gamma\left({2\alpha\over{b}}\right)}},\\
C^{R}(\alpha) &=&{\tilde\mu}\pi\gamma\left(1+b^{-2}\right)
{\Gamma\left(1+{2\alpha\over{b}}-{1\over{b^2}}\right)
\Gamma\left(-{2\alpha\over{b}}\right)\over{
\Gamma\left(-{2\alpha\over{b}}+{1\over{b^2}}\right)
\Gamma\left(1+{2\alpha\over{b}}\right)}},
\eeaq
where ${\tilde\mu}$, the cosmological constant of the dual theory,
has been related to that of the $N=2$ SLFT in \cite{AKRS}. 

These two-point functions can be expressed as
\beaq
G^{NS}_{\alpha}(\xi,\xi')
&=&U^{R}\left(\alpha-{b\over{2}}\right)
{\cal G}^{NS}_{+}(\xi,\xi')+C^{NS}(\alpha)U^{R}
\left(\alpha+{b\over{2}}\right) {\cal G}^{NS}_{-}(\xi,\xi')\\
G^{R}_{\alpha}(\xi,\xi')
&=&U^{NS}\left(\alpha-{b\over{2}}\right)
{\cal G}^{R}_{+}(\xi,\xi')+C^{R}(\alpha)U^{NS}
\left(\alpha+{b\over{2}}\right) {\cal G}^{R}_{-}(\xi,\xi')
\eeaq
where ${\cal G}_{\pm}(\xi,\xi')$'s are expressed in terms of the special
conformal blocks
\beq
{\cal G}^{NS}_{\pm}(\xi,\xi')={|\xi'-{\overline \xi'}|^{2\Delta^{NS}_{\alpha}
-2\Delta^{R}_{-b/2}}\over{|\xi-{\overline \xi'}|^{4\Delta^{NS}_{\alpha}}}}
{\cal F}^{NS}_{\pm}(\eta),\quad 
{\cal G}^{R}_{\pm}(\xi,\xi')={|\xi'-{\overline \xi'}|^{2\Delta^{R}_{\alpha}
-2\Delta^{NS}_{-b/2}}\over{|\xi-{\overline \xi'}|^{4\Delta^{R}_{\alpha}}}}
{\cal F}^{R}_{\pm}(\eta),
\nonumber
\eeq
with
\beq
\eta={(\xi-\xi')({\overline \xi}-{\overline \xi'})\over{(\xi-{\overline \xi'})
({\overline \xi}-\xi')}}.
\nonumber
\eeq
Here, the conformal blocks are given by the hypergeometric functions
\bea
{\cal F}^{NS}_{+}(\eta)&=&\eta^{{\alpha \over{b}}}
(1-\eta)^{-{1\over{b^2}}-{1\over{2}}}
F\left({2\alpha\over{b}},1+{1\over{b^2}};{2\alpha\over{b}}
-{1\over{b^2}}+ 1;\eta\right)\\
{\cal F}^{NS}_{-}(\eta)&=&\eta^{{\alpha \over{b}}}
(1-\eta)^{-{2\over{b^2}}+1}
F\left({1\over{b^2}},1-{2\alpha\over{b}}+{2\over{b^2}};-{2\alpha\over{b}}
+{1\over{b^2}}+ 1;\eta\right)\\
{\cal F}^{R}_{+}(\eta)&=&\eta^{{\alpha\over{b}}-{1\over{2}}}
(1-\eta)^{-{1\over{b^2}}-{1\over{2}}}
F\left({2\alpha\over{b}}+1,{1\over{b^2}};{2\alpha\over{b}}
-{1\over{b^2}}+2;\eta\right)\\
{\cal F}^{R}_{-}(\eta)&=&\eta^{{\alpha \over{b}}-{1\over{2}}}
(1-\eta)^{-{2\over{b^2}}+{1\over{2}}}
F\left({1\over{b^2}},-1-{2\alpha\over{b}}+{2\over{b^2}};-{2\alpha\over{b}}
+{1\over{b^2}};\eta\right).
\eea

On the other hand, one can compute the two-point functions as
both $R^{+}_{-1/2b}$ and $N_{\alpha}$ or $R^{-}_{\alpha}$ 
approach on the boundary. 
The fusion of the degenerate operator with the boundary is described
by a special bulk-boundary structure constant which could be computed
as a boundary screening integral with one insertion of the boundary 
interaction of the dual $N=2$ theory if it were known.
Since we can not fix it, we denote the unknown constant just as
${\cal R}(-1/2b)$.
Then, we can obtain the system of functional relations as follows:
\beaq
{\cal R}\left(-{1\over{2b}}\right)
U^{NS}(\alpha)&=&{\Gamma(1-{1\over{b^2}}+{2\alpha\over b})
\Gamma(-{2\over b^2})\over \Gamma(1-{2\over b^2}+{2\alpha\over b})
\Gamma(1-{1\over b^2})} U^R\left(\alpha-{1\over 2b}\right)\nonumber\\
&+&C^{NS}(\alpha){\Gamma(1+{1\over b^2}-{2\alpha\over b})\Gamma(-{2\over b^2})
\over \Gamma(1-{2\alpha\over b})\Gamma(1-{1\over b^2})}
U^R\left(\alpha+{1\over 2b}\right)\\
{\cal R}\left(-{1\over{2b}}\right)
U^R(\alpha) &=&{\Gamma({2\alpha\over b}-{1\over b^2})
\Gamma(-{2\over b^2})\over \Gamma ({2\alpha\over b}-{2\over b^2})
\Gamma (1-{1\over b^2})}U^{NS}\left(\alpha-{1\over 2b}\right)\nonumber\\
&+&C^{R}(\alpha){\Gamma({1\over b^2}-{2\alpha\over b})\Gamma(-{2\over b^2})
\over \Gamma ({-2\alpha\over b})\Gamma(1-{1\over b^2})}
U^{NS}\left(\alpha+{1\over 2b}\right).
\eeaq
Although we do not know the bulk-boundary structure constant,
we can elliminate it by taking ratio of above equations and 
find one relation which is complitely fixed.
It can be shown that the one-point functions Eqs.(\ref{oneptns})
and (\ref{oneptr}) indeed satisfy this relation.
This means not only that the one-point functions obtained from the modular
bootstrap procedures are consistent with the $N=2$ SLFT actions,
but also that the $N=2$ theory proposed in \cite{AKRS} is indeed 
dual to the $N=2$ SLFT.
Furthermore, we can find the bulk-boundary structure constant as follows:
\beq
{{\cal R}\left(-{1\over{2b}}\right)\Gamma\left(1-{1\over{b^2}}\right)
\over{\Gamma\left(-{2\over{b^2}}\right)
\sqrt{{\tilde\mu}\pi\gamma\left(1+{1\over{b^2}}\right)}}}
=\cosh\left({2\pi s\over{b}}\right).
\eeq
Along with Eq.(\ref{boundparam}),
this equation relates the boundary cosmological constant of 
the $N=2$ SLFT with that of the dual $N=2$ theory. 

\subsection{Semiclassical Checks}

These results can be checked, semiclassically.
As $b\to 0$, the $N=2$ SLFT (\ref{N2L}) can be described by the 
Schr\"odinger equation 
\beq
\left[-{1\over{2}}{\partial^2\over{\partial\Phi_0^2}}
+{\pi^2\mu^2b^2\over{8}} e^{2b\Phi_0}\right]
\Psi_P(\Phi_0)=2P^2\Psi_P(\Phi_0)
\eeq
where $\Phi_0=(\phi^{+}_{0}+\phi^{-}_{0})/2$ in terms of
the zero-modes of $\phi^{\pm}$.
Solving this, one can derive the reflection amplitude
\beq
D^{NS}(P,\omega)=-\left({\pi^2\mu^2\over{16}}\right)^{-2iP/b}
{\Ga\left(1+{2iP\over{b}}\right)\over{\Ga\left(1-{2iP\over{b}}\right)}},
\eeq
and can show that this is consistent with Eq.(\ref{genrefns}). 

Another interesting check is to compute the inner product
semiclassically
\beq
\langle ({\mathbf s})\vert [P,\omega]\rangle
=\int_{-\infty}^{\infty}d\Phi_0\Psi_{B_{s}}(\Phi_0)\Psi_P(\Phi_0)
\eeq
where the boundary state can be expressed by the boundary Lagrangian
following \cite{FZZ}
\beq
\Psi_{B_{s}}(\Phi_0)=\exp\left(-8\pi^2b^{-1}{\overline\mu_B}^2 
e^{b\Phi_0}\right),
\eeq
with ${\overline\mu_B}$ a boundary cosmological constant defined 
in Eq.(\ref{bcosmo}).
From this, one can find
\beq
\langle ({\mathbf s})\vert [P,\omega]\rangle
=\left(\frac{8\pi^2\overline\mu_B^2}{b}\right)^{-2iP/b}
\Ga\left({2iP\over{b}}\right),
\eeq
which, along with Eq.(\ref{boundparam}), agrees with Eq.(\ref{ampns}).

\section{Discussions}

Using the one-point function, one can find a density of states
which can be related to the boundary two-point functions.
The partition function $Z^{NS}_{s, s'}(q,y,t)$
with contiuous BCs on both boundaries parametrized by $s$ and $s'$
can be obtained as
\beq
Z^{NS}_{s, s'}(q,y,t)
=\int_{-\infty}^{\infty} dP\int_{-\infty}^{\infty}d\omega
\chi^{NS}_{[P,\omega]}(q',y',t')
\Psi^{NS}_{\mathbf s} (P){\Psi^{NS}_{\mathbf s'}}^{\dagger}(P,\omega)
\eeq
with the amplitude (\ref{ampns}). 
This can be rewritten as 
\beaq
Z^{NS}_{s, s'}(\tau)
&=&b\int_{-\infty}^{\infty} dP' d\omega' dP d\omega
e^{- 4 i \pi P P'}e^{-i\pi b^2\omega\omega'}
\chi^{NS}_{[P,\omega]}(q,y,t)
\Psi^{NS}_{\mathbf s}(P',\omega')
{\Psi^{NS}_{\mathbf s'}}^{\dagger}(P',\omega') \nonumber\\
&=& \int_{0}^{\infty} dP\int_{0}^{\infty} d\omega
\chi^{NS}_{[P,\omega]}(q,y,t)
\rho^{NS}_{ss'}(P,\omega),
\eeaq
where $\rho^{NS}_{ss'}(P,\omega)$ is the density of states,
\beq
\rho^{NS}_{ss'}(P,\omega)={4\over{\pi^2}}\int_{-\infty}^{\infty}du
\int_{-\infty}^{\infty}dv 
\left[{\cosh(bv)+\cos(u)\over{2\sinh(bv)\sinh(v/b)}}\right]
\cos(2sv)\cos(2s'v) e^{-2ivP-iu\omega}.
\eeq
This quantity is not well-defined at $P=0$
and is to be properly regularized.  
This density of states is, on the other hand, conjectured to be
related with the two-point function $d_B^{NS}(P,\omega|s,s')$ of
boundary operator $n_{\alpha{\overline\alpha}}^{ss'}$ by
\beq
\rho^{NS}_{s,s'}(P,\omega) = - {i \over 2\pi } 
{d \over dP} \log d^{NS}_B (P,\omega|s,s'). 
\eeq

In this paper, we have derived one-point functions of the $N=2$ SLFT
and provided various consistency checks.
Our consistency checks also confirms the validity of the dual
$N=2$ super-CFT conjecture in \cite{AKRS} and the boundary action
proposed in \cite{AY}.
It would be interesting to provide the boundary action for the 
dual theory so that one can complete the boundary bootstrap 
procedure for the $N=2$ SLFT with the boundary.
Our result can be applied to 2d superstring theories and related 
topics generalizing the work on the $N=1$ SLFT \cite{DKKMMS}.
These results also can be used to study the integrable quantum field theories
with $N=2$ supersymmetry which can be constructed as perturbed 
$N=2$ SLFT.
We hope to report a progress in this direction in future publications.

\section*{Note Added}

After finishing this article, we found a paper \cite{EguSug} 
which results overlap with ours in large part.
While this paper deals mainly with the vacuum BC and
applications and implications to the string theory,
we are more interested in the one-point functions for a continuous
BC in terms of the boundary cosmological constant
and their relations to the conformal bootstrap equations.

\section*{Acknowledgments}

We thank Jaemo Park for helpful discussions.
This work was supported in part by Korea Research Foundation 
2002-070-C00025. 
MS is supported by the Brain Pool program from Korean Association 
Science and Technology.

%\appendix

\end{document}